\documentclass[prl,twocolumn,showpacs,preprintnumbers,amsmath,amssymb,]{revtex4-1}

\usepackage{comment}
\usepackage{graphicx}
\usepackage{dcolumn}
\usepackage{bm}
\usepackage{float}
\usepackage{color}
\usepackage{soul,xcolor}
\setstcolor{red}

\newcommand{\red}[1]{\textcolor{black}{#1}}
\newcommand{\blue}[1]{\textcolor{black}{#1}}
\newcommand{\textcomment}[1]{\ignorespaces}

\usepackage{amsmath}
\usepackage{braket} 

\begin{document}

\preprint{APS/123-QED}

\title{Observation of Quantum Interference and Coherent Control in a Photochemical Reaction}

\author{David B. Blasing,$^1$ Jes\'{u}s P\'{e}rez-R\'{\i}os,$^2$, Yangqian Yan,$^1$ Sourav Dutta,$^{1,3}$, Chuan-Hsun Li,$^4$ Qi Zhou,$^{1,5}$ and Yong P. Chen$^{1,4,5,}$}
\email[ ]{yongchen@purdue.edu} 
\affiliation{
$^1$Department of Physics and Astronomy, Purdue University, West Lafayette, IN  47907\\
$^2$School of Materials Sciences and Technology, Universidad del Turabo, Gurabo, Puerto Rico 00778\\
$^3$Department of Physics, Indian Institute of Science Education and Research, Bhopal 462066, India\\
$^4$School of Electrical and Computer Engineering, Purdue University, West Lafayette, IN  47907\\
$^5$Purdue Quantum Center, Purdue University, West Lafayette, IN  47907
}



\date{\today}


\begin{abstract}
Coherent control of reactants remains a longstanding challenge in quantum chemistry. In particular, we have studied laser-induced molecular formation (photoassociation) in a Raman-dressed spin-orbit-coupled $^{87}$Rb Bose-Einstein condensate, whose spin quantum state is a superposition of multiple bare spin components. In contrast to the notably different photoassociation-induced fractional atom losses observed for the bare spin components of a statistical mixture, a superposition state with a comparable spin composition displays the same fractional loss on every spin component. We interpret this as the superposition state itself undergoing photoassociation. For superposition states induced by a large Raman coupling and zero Raman detuning, we observe a nearly complete suppression of the photoassociation rate. This suppression is consistent with a model based upon quantum destructive interference between two photoassociation pathways for colliding atoms with different spin combinations. This model also explains the measured dependence of the photoassociation rate on the Raman detuning at a moderate Raman coupling. Our work thus suggests that preparing atoms in quantum superpositions may represent a powerful new technique to coherently control photochemical reactions.
\end{abstract}

\maketitle
	



 
Quantum coherent control of atomic processes has been a significant triumph of atomic, molecular, and optical physics. Extending such coherent control to molecular processes is an active and interesting research area. In particular, the study of coherent control of photochemical molecular processes has focused on light-based control or control of the initial and final quantum states (for reviews see Refs. \cite{Shapirobook,Brumer1992,Koch2012,Brumer2012}). Theoretical studies have concerned both the manipulation of light parameters, such as the pulse trains, polarization, relative phases, etc.,  \cite{Tannor1985,Tannor1986,Vrakking1997,Hild2016,Schlawin2017,Kallush2017,JPR2017,Lima2017} and the initial or final quantum states \cite{Brumer1986}. Experimentally, tailored light pulses have been shown to control isomerization, photoassociation (PA), and photodissociation ~\cite{iso,Levis2001,Lozovoy2008,Carini2015,Carini2016,Paul2016,Endo2017,Levin2015}. However, there is much lesser experimental study of influencing molecular processes by coherently controlling the reactants. Such a difficulty can arise from incoherent population in many scattering states due to finite experimental temperatures or an incomplete understanding of the quantum molecular processes.
 
\begin{figure}[hbt]
\centering
\includegraphics[width=8.5cm,trim=0cm 0cm 0cm 0cm,clip=true]{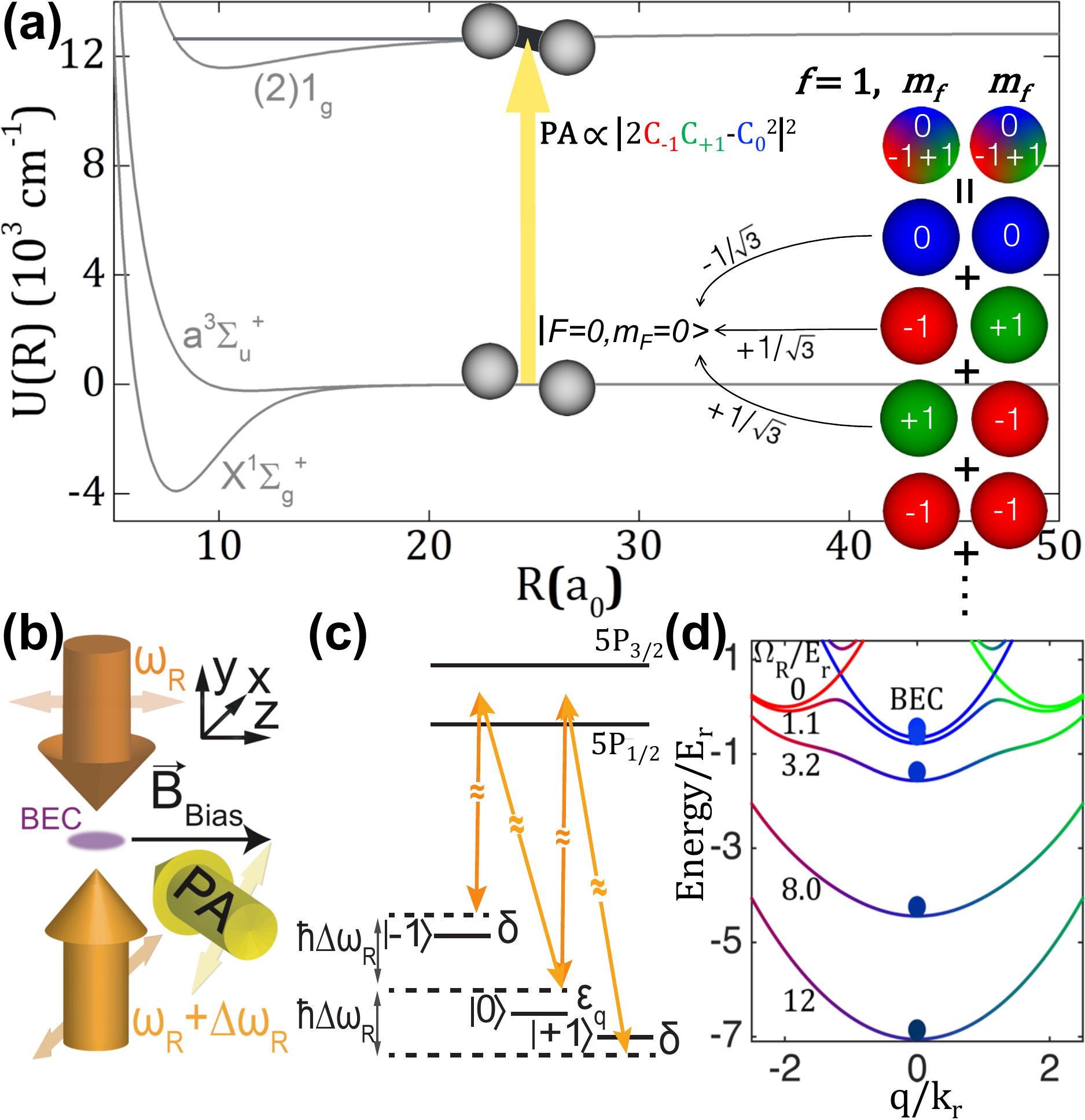}
\caption{\red{Energy level diagrams and experimental setup.} (a) Relevant molecular potential energy curves. Depicted at the right is a scattering state of a pair of atoms (the tri-colored spheres) whose spin part of their quantum state is a superposition of different bare $m_f$ spin states. Beneath is the decomposition of the scattering state as that of various pairs of atoms (the mono-colored spheres) with bare $m_f$ spin states. The superposition coefficients are denoted $C_{-1},C_{0},$ and $C_{+1}$; red, blue, and green represent $m_f=$ -1, 0, and +1, respectively. The non-zero CG coefficients for the $\ket{F=0, m_F=0}$ component of the various pairs of bare spins in the scattering state are shown near the thin black arrows. (b) Laser geometry. Two Raman lasers with angular frequencies $\omega_R+\Delta\omega_R$ and $\omega_R$ propagate along $\pm\hat{y}$ and have linear polarizations along $\hat{x}$ and $\hat{z}$. The frequency difference between the Raman beams $\Delta\omega_R/2\pi$ is $ 3.5$ MHz. A PA laser propagates in the x-z plane (due to spatial constraints) and has a linear polarization with components along all three axes. (c) Atomic energy diagram. A Zeeman bias magnetic field, $|\vec{B}_\mathrm{Bias}|\approx 5$ G is used to tune the Raman detuning $\delta.$ (d) Dressed band structures for Raman coupling $\Omega_R= 0,\,1.1,\,3.2, 8.0$ and $12\,E_r$, all with $\delta=0$. Dots represent BECs adiabatically loaded to the band minima at $q=0$. Panels (a) to (c) are not to scale.}
\label{fig:Exp_dgm}
\end{figure}

In this work, we explore the question: what happens in a chemical reaction if the reactants are prepared in quantum superposition states? Here we report spin-dependent PA experiments using a $^{87}$Rb Bose-Einstein condensate (BEC). Photoassociation \cite{Jones2006} is a light-assisted chemical process where two atoms absorb a photon while scattering, and bind into an excited molecular state. Our Bose-Einstein condensates are at ultracold temperatures and populate only a small number of scattering channels. \red{In our experiment we have prepared atoms in spin-momentum quantum superposition states, and a pair of such atoms can couple to an excited molecular state simultaneously through two atomic scattering channels. These two channels of different spin combinations both contribute to the coupling, but with opposite sign due to opposite Clebsch-Gordan (CG) coefficients. The relative amplitudes of the two contributions depend on the superposition state. \red{The spin portion of a representative scattering state is shown in Fig. \ref{fig:Exp_dgm} (a), along with the relevant molecular potential energy curves plotted against the internuclear separation $R$ in units of Bohr radius $a_0$ \cite{Allouche2012}.} Our system exploits the intrinsic quantum nature of PA and our tunable superposition states of the reactant atoms allows us to observe a nearly total suppression of the molecular formation, thus representing a significant step forward for coherent chemistry.} 


Our experiment begins with a $^{87}$Rb BEC of $\sim1.5\times10^4$ atoms in the $f=1$ hyperfine state,  \red{which is produced via all optical evaporation in a cross-beam optical dipole trap created by a 1550 nm laser. The trap has a characteristic trap frequency $\bar{\omega}=(\omega_x\omega_y\omega_z)^{1/3}\sim2\pi\times(140\times140\times37)^{1/3}\,\,\text{Hz}=2\pi\times(90$ Hz) \cite{Olson2013}. Tuning the magnetic field during the evaporation can lead to a BEC with bare $m_f=-1,0,+1$ }spin state, or a statistical mixture of all three. 

After preparing a bare BEC in spin state $m_f=0$, we can load the BEC into a spin-momentum superposition \blue{by adiabatically applying a pair of counter-propagating Raman beams with wavelengths near 790.17 nm, see Fig. \ref{fig:Exp_dgm} (b). The Raman beams couple the $m_f$ states, as shown in Fig. \ref{fig:Exp_dgm} (c), and ``dress" the atoms into superpositions of the $m_f$ spin states and mechanical momenta: $\sum\limits_{i=1}^3 C_i\ket{m_{f},p}_i=C_{-1}\ket{-1,\hbar(q+2k_r)}+C_{0}\ket{0, \hbar q}+C_{+1}\ket{+1,\hbar(q-2k_r)}$, where $p$ and $\hbar q$ are the mechanical momentum and quasimomentum along $\hat{y}$ respectively, $\hbar k_r$ is the single photon recoil momentum of the Raman lasers with wavelengths near 790.17 nm, and $\hbar$ is the reduced Planck's constant. The quasimomentum is the canonical momentum conjugate of the position along $y$. For more details of our setup, see Refs. \cite{Olson2014,Olson2017}. Our Raman coupling scheme is largely similar to previous works \cite{Lin2009,Lin2011}. The Hamiltonian for the Raman light-atom interaction is}

\begin{equation}
	\mathcal{H}
	=\begin{pmatrix}
\frac{\hbar^2}{2m}(q+2k_r)^2-\delta & \frac{\Omega_R}{2} & 0 \\ 
\frac{\Omega_R}{2}  & \frac{\hbar^2}{2m}q^2-\epsilon_q & \frac{\Omega_R}{2} \\ 
0 & \frac{\Omega_R}{2}  & \frac{\hbar^2}{2m}(q-2k_r)^2+\delta \\ 
	\end{pmatrix},
	\label{eq:ham_countprop}
\end{equation}

\noindent where $m$ is the $^{87}$Rb mass, $\delta$ is the Raman detuning\textcomment{(\red{inferred from the two Raman-Rabi resonances at $\delta=\pm4.65\,\,E_r$)}}, $\Omega_R$ is the Raman coupling (calculated from the measured frequency of resonant Raman-Rabi oscillations), $E_r=\hbar^2 k_r^2/2m=h\times(3.68$ kHz) is the recoil energy, and $\epsilon_q=0.65\,E_r$ is the quadratic Zeeman shift. Unless otherwise stated, $\Omega_R$ and $\delta$ carry uncertainties of 10\% and $\pm\,0.5\,E_r$ respectively. \red{The eigenstates of Eq. \ref{eq:ham_countprop} are the ``dressed" spin-momentum superposition states described above.} The $q$-dependent eigen-energies of the atoms form the ``dressed" bands, with a few representative examples shown in Fig. \ref{fig:Exp_dgm}    (d). Red, blue, and green colors reflecting the proportion of $m_f=-1,0,$ and $+1$ in the spin-momentum superposition state. The small dots represent the BECs adiabatically prepared at the band minimum. Such Raman-coupling has been used previously to induce synthetic partial waves \cite{Williams2012}, modify an s-wave Feshbach resonance \cite{Williams2013}, and couple singlet and triplet scattering states \cite{Fu2013}. 

After preparing our BEC in a statistical mixture or a spin-momentum superposition state, we apply a PA laser with wavelength $781.70$ nm and typical intensity $I_{PA}$ of a few $\text{W}/\text{cm}^2$ for a time $t_{PA}$ of a few ms \footnote{The beam waist of both the PA and the Raman lasers at the BEC location is about an order of magnitude larger than the in-situ BEC size.}. The frequency of the PA laser is tuned to the $(2)1_g$ excited molecular state with vibrational quantum number $152$ (see again Fig. 1 (a)), and corresponds to the PA line $\epsilon$ in the Fig. 1 of Ref. \cite{Hamley2009}.  After the PA pulse, the PA, Raman (in the case of the superposition state), and the dipole lasers are simultaneously switched off to allow the BEC to undergo 15 ms of time-of-flight expansion (TOF). During the initial portion of this expansion, we use a Stern-Gerlach magnetic field gradient to separate atoms in the different $m_f$ spin states. At the end of the TOF expansion, we apply absorption-based imaging to extract the atom numbers in different spin-momentum projections. Then, this sequence is repeated at various PA frequencies to obtain a PA spectrum.
 
To extract the PA rate constant, $k_{PA}$, due to a PA process, we follow a procedure similar to those in Refs. \cite{McKenzie2002,Theis2004}. The two body rate equation describing the time-dependent BEC density \red{of the atoms participating in PA}, ${\rho}(t,\vec{r})$, is $d{\rho}(t,\vec{r})/dt = -k_\mathrm{PA}\rho^2(t,\vec{r})$. In our experiment, $\Gamma_{stim}<<\Gamma_{spon}$ (where $\Gamma_{stim}$ and $\Gamma_{spon}$ are the stimulated and spontaneous emission rates, respectively), and $k_\mathrm{PA}$ has a Lorenztian dependence with respect to the PA frequency \cite{Bohn1997}. Solving for $\rho(t,\vec{r})$ and spatially integrating it over a Thomas-Fermi BEC density profile yields an expression for the atom number, $N(\eta)$, remaining after a PA pulse,

\begin{figure}[htb]
\centering
\includegraphics[width=8.5cm,trim=.5cm 2.1cm 5.3cm 0cm,clip=true]{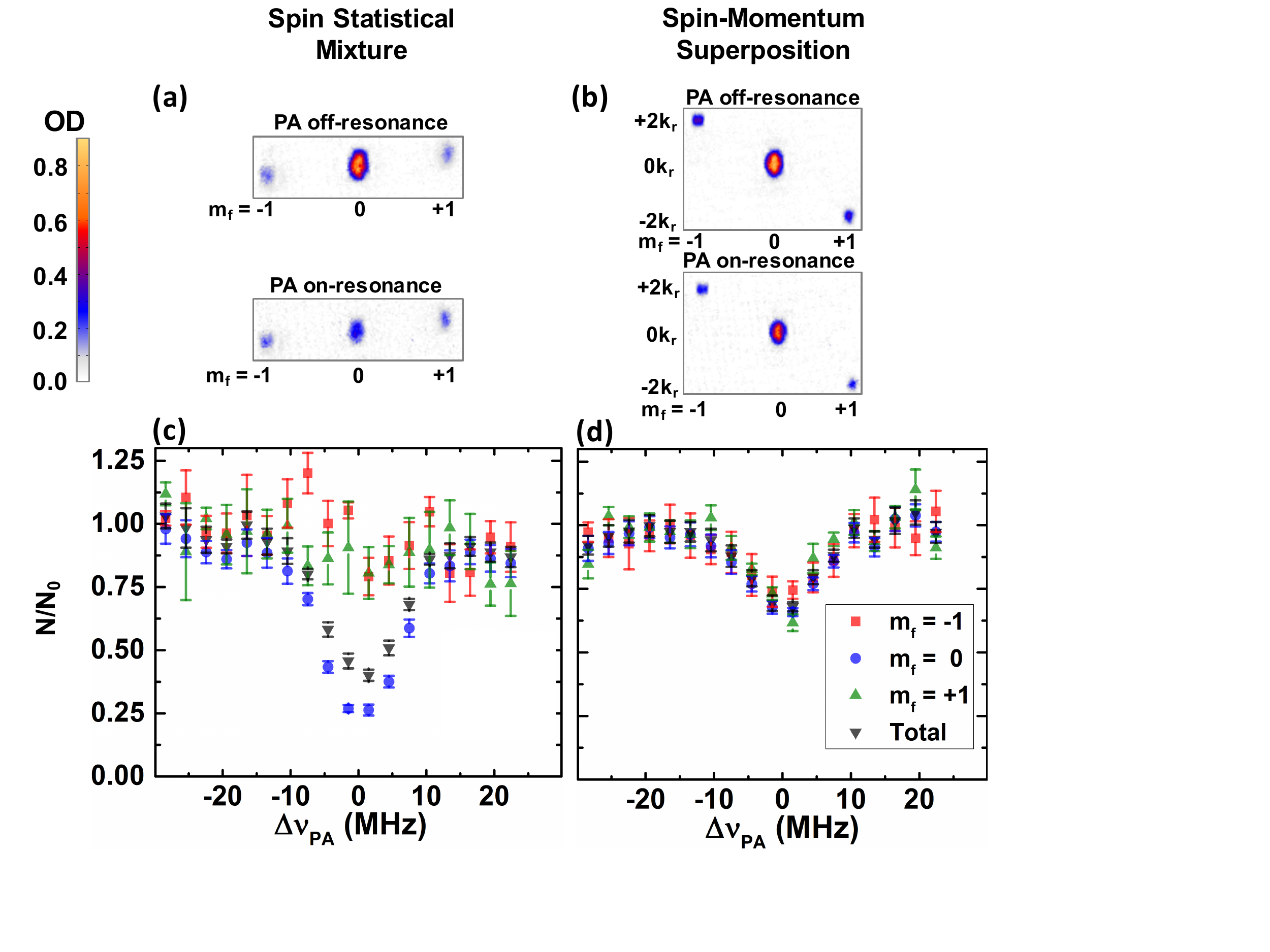}
\caption{Photoassociation of BECs with atoms in a spin statistical mixture (with Raman coupling $\Omega_{R}=0\,\,E_{r}$) versus a spin-momentum superposition state with similar atom number and spin composition ($\Omega_{R}=8.0\,\,E_{r}$ and Raman detuning $\delta=0.0\,\,E_{r}$). (a) and (b): the average optical density (OD) images with PA off and on resonance for the spin statistical mixture (a) and the spin-momentum superposition state (b). (c) and (d): the extracted normalized atom numbers of the $m_f$ components and the total of BECs at various PA detunings ($\Delta\nu_\text{PA}$) from resonance for the spin statistical mixture (c) and spin-momentum superposition (d). The atom numbers of every $m_f$ component or the total are normalized by the corresponding fitted values of the off-resonant atom number $N_0$ and the error bars are the standard error of the mean. Both the OD images and data points are averages of 5 to 7 experimental runs.}
\label{fig:PA_spin_mom_sup_norm}
\end{figure}

\begin{equation}
\begin{split}
N(\eta) =  & N_0\frac{15}{2}\eta^{-5/2}[\eta^{1/2}+\frac{1}{3}\eta^{3/2}- \\
& (1+ \eta)^{1/2}\mathrm{tanh}^{-1}(\sqrt{\eta/(1+\eta)})],
\label{eq:eta}
\end{split}
\end{equation}

\noindent where $\eta=k_\mathrm{PA}\rho_0t_{PA}$ is a dimensionless parameter indicating the strength of the PA pulse, $\rho_0$ is the peak atomic density at the center of the BEC, and $N_0$ is the off-resonant atom number with no PA loss. We denote the resonant $k_\mathrm{PA}$ for a BEC composed of $m_f=0$ bare spin (or spin-momentum superposition) states as $k_{0,0}$ (or $k_{sup}$). 

First, as shown in Fig. \ref{fig:PA_spin_mom_sup_norm}, we compared PA of BECs  in a spin statistical mixture to that of BECs in a spin-momentum superposition state, with a nearly identical atom number and spin composition in the two cases. For the spin-momentum superposition state, we used $\Omega_R=8.0\,E_r$ and $\delta=0\,E_r$ (see also Fig. \ref{fig:Exp_dgm} (d)). In panels (a) and (b), we show the optical density (OD) images for PA both on and off-resonance for the spin statistical mixture (a) and the spin-momentum superposition state (b). For the spin statistical mixture, the PA-induced loss for the $m_f=0$ component is notably larger than that for the $m_f=\pm1$ components. \red{The lower reduction of the OD was due to the lower $\rho_0$ for the $m_f=\pm\,1$ components and that each molecule formed by the PA process reduced the $m_f=0$ atom number by two, but for $m_f=\pm1$, only by one each.} However, for PA on a spin-momentum superposition state, we observed comparable PA-induced loss among all three $m_f$ components. In panels (c) and (d), for each $m_f$ component and the total, we show the normalized atom number ($N/N_0$) at various PA detunings ($\Delta\nu_{PA}$) from the resonance for the statistical mixture (c) and the spin-momentum superposition state (d). Each PA spectrum for every $m_f$ component or the total was fitted to Eq. \ref{eq:eta} to extract the appropriate $N_0$ and then normalized. \red{The $N_0$ were $\sim$ (1.2,7.0,1.1)$\times10^3$ and (1.5,6.9,1.3)$\times10^3$ for the $(m_f=-1,0,+1)$ components of the statistical mixture and superposition state respectively.} For the spin statistical mixture, $(79\pm2)$\% of the $m_f=0$ atoms were lost on resonance, but less than $\sim25\%$ were lost for the $m_f=\pm1$ components. For the dressed BECs, all $m_f$ components lost $(36\pm2)$\%. All the data were taken using PA pulses with identical parameters. 

\begin{figure}[hbt]
\centering
  \includegraphics[width=8.5cm,trim= 0cm 0cm 0cm 0cm,clip=true]{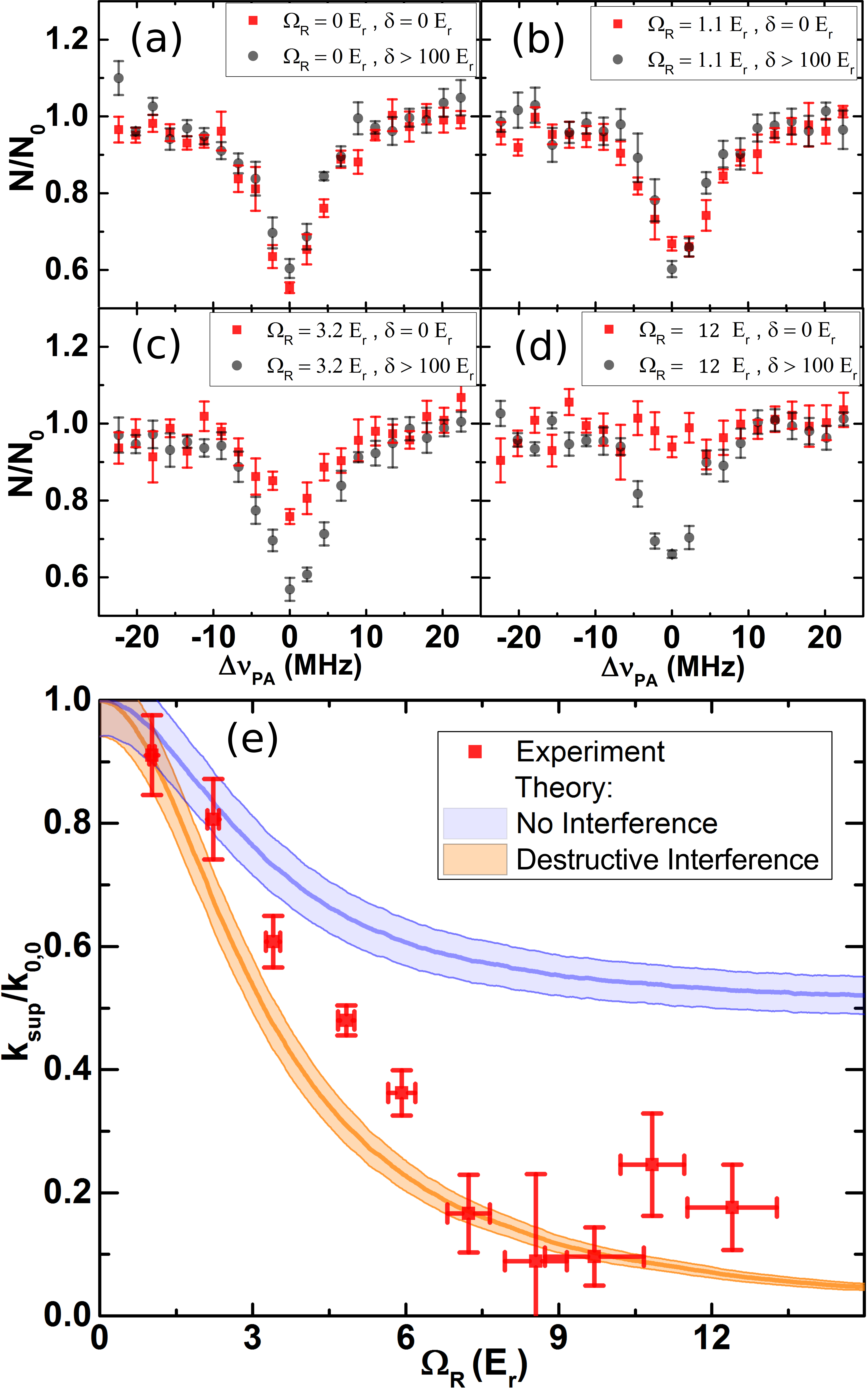}
  \caption{Photoassociation of atoms in spin-momentum superpositions at various values of the Raman coupling $\Omega_R$. Panels (a)-(d), normalized atom loss for BECs of atoms in spin-momentum superpositions (or spin state ${m}_f=0$) with Raman detuning $\delta=0\pm0.5\,E_r\,\,($or $\delta>100\,E_r$) are shown as red squares (or black circles). The values of $\Omega_R/E_r$ were: 0 (a), 1.1 (b), 3.2 (c), and 12 (d). Error bars are the standard error of the mean of 6 experimental runs. Panel (e), normalized photoassociation rate, $k_{sup}/k_{0,0}$, for BECs at various $\Omega_\mathrm{R}$ with $\delta=0\,\,E_{r}$. The orange (blue) bands are theoretical predictions with (without) the destructive interference term in Eq. \ref{eq:spinsquare}. The band boundaries reflect one standard deviation of the predicted values given our experimental uncertainties.}
\label{fig:fraclossfig}
\end{figure}

To further explore this phenomenon, we prepared BECs with atoms in several spin-momentum superposition (or $m_f=0$ spin) states by using $\Omega_R=0,\,1.1\,,3.2$, and $12\,\,\text{E}_r$ with $\delta=0\,E_r$ (or $\delta\sim100\,E_r$) and plotted the normalized PA spectra in Fig. \ref{fig:fraclossfig} (a)-(d) using red squares (or black circles)\textcomment{The eight normalized PA spectra were obtained as follows. For each of the eight combinations of $\Omega_R$ and $\delta$, we took six individual scans which were averaged to form eight unnormalized PA spectra. Then, these eight unnormalized curves were fitted with Eq. 2, normalized to their $N_0$, and plotted. The spectra were taken in a non-linear and alternating fashion to help address any systematic drifts while using the same PA pulse ($t_{PA}=3.2$ ms and $I_{PA}=6.0\pm$ $ 0.7 \,\,\text{W}/\text{cm}^2$). There is small, but present, reduction in the $m_f=0$ loss with increasing $\Omega_R$. We attribute this to a slight misalignment of the BEC induced by the Raman beams at high power. This small systematic effect is why we measure the bare $m_f=0$ loss with $\Omega_R\neq0$ and large $\delta$, instead of with $\Omega_R=\delta=0$. We observed no systematic effect of the on $k_{0,0}$ at different $\delta$.}. With $\delta\sim100\,E_r$, the Raman beams did not dress the atoms into spin-momentum superposition states, and these BECs therefore remained in the $m_f=0$ bare spin state and displayed comparable loss $(\sim40\%)$ for all $\Omega_R$. However, the loss for BECs in spin-momentum superpositions decreases with increasing $\Omega_R$. At $\Omega_R=12\,E_r$, no loss is apparent. For all the data in panels (a) to (d), we used comparable total off-resonant BEC atom numbers ($N_0\sim(1.1\pm0.1)\times10^4$) and square PA pulses with $t_{PA}=3.2$ ms and $I_{PA}=6.0\pm0.7 \,\,\text{W}/cm^2$. 

We also fitted photoassociation spectra measured \textcomment{For each measurement of $k_{sup}/k_{0,0}$, we took eight PA spectra. These sets of eight were repeated about thirty times at various $\Omega_R$. Afterwards, they were binned, averaged and plotted. The PA spectra for $\delta=0$ and $\delta\sim100\,E_r$ were both fitted to extract their values of $\eta$. Then, since $\rho_0$ and the PA pulse time were independently measured, we then calculated $k_{sup}$ and $k_{0,0}$.} with $\delta=0$ (or $\delta\sim100\,E_r$) to Eq. \ref{eq:eta} and extracted $k_{sup}$ (or $k_{0,0}$), and then plotted $k_{sup}/k_{0,0}$ in Fig. \ref{fig:fraclossfig} (e). We used PA pulses with $I_{PA}=6.1\pm0.7\,\,\text{W}/\text{cm}^2$ and $t_{PA}$ of 2 to 8 $ms$ to induce a repeatable but unsaturated loss of (10 to 40)\%. Also included are solid bands, which are predictions for $k_{sup}/k_{0,0}$ derived as follows. \red{The molecular hyperfine state excited by our chosen PA transition has total angular momentum $F=1$ and nuclear spin $I=1,$ and only couples to a pair of colliding atoms (represented by subscripts a and b in the following) whose total angular momentum $\ket{F=f_a+f_b,m_F=m_{f,a}+m_{f,b}} = \ket{0,0}$. Using the single particle basis, $\ket{f_a,m_{f,a}}\ket{f_b,m_{f,b}}$, $\ket{F=0,m_F=0}=(\ket{1,+1}\ket{1,-1}+\ket{1,-1}\ket{1,+1}-\ket{1,0}\ket{1,0})/\sqrt{3}$. Thus, there are two allowed pathways for the PA transition (after accommodating the indistinguishability of bosons): one in which both atoms have $m_{f}=0$ and another with atoms in $m_{f}=\pm1$. Both pathways contribute to this PA process with opposite signs due to the opposite Clebsch-Gordan (CG) coefficients ($\pm1/\sqrt{3}$)\textcomment{Two $f=1$ atoms can have $F=2$ components, but these are inactive for our particular PA process.}.} We note $k_{PA}\propto|\bra{\psi_{mol}}\vec{d}\cdot\vec{E}\ket{\psi_{scat}}|^2$ with a spin independent proportionality factor \footnote{See the Supplemental Information.}, where $\ket{\psi_{scat}}$ and $\ket{\psi_{mol}}$ are the total wavefunctions for the scattering state and molecular state respectively, $\vec{E}$ is the electric field of the PA laser, and $\vec{d}$ is the dipole operator. The spin portion of $\ket{\psi_{scat}}$ for two Raman dressed atoms (labeled by subscripts $a$ and $b$) is $\sum\limits_{i=-1}^{+1}\sum\limits_{j=-1}^{+1}C_iC_j\ket{f=1,m_{f}=i}_a\otimes\ket{f=1,m_{f}=j}_b$. Using the CG coefficients, the probability amplitude of the $\ket{F=0,m_F=0}$ component of the scattering wavefunction is therefore $(2C_{-1}C_{+1}-C_0^2)/\sqrt{3}$. We arrive at \footnote{See the Supplemental Information. In writing Eq. \ref{eq:spinsquare}, we have ignored in the Franck-Condon factor the relative momentum imparted from the Raman beams. For the purpose of our PA resonance, this is justifiable since the length scale set by the \red{wavelength of the Raman beams is $\sim 1\,\mu$m}, but the length scale of the PA process is significantly shorter, less than 2 nm.}

\begin{eqnarray}
\label{eq:spinsquare}
k_{sup}/k_{0,0}=|C_0^2|^2+4|C_{-1}C_{+1}|^2-4\text{Re}[C_0^2C^*_{-1}C^*_{+1}].\,\,\,\,\,\,\,
\end{eqnarray}

\begin{figure}[htb]
\centering
\includegraphics[width=8.5cm,trim=0cm 0cm 0cm 0cm,clip=true]{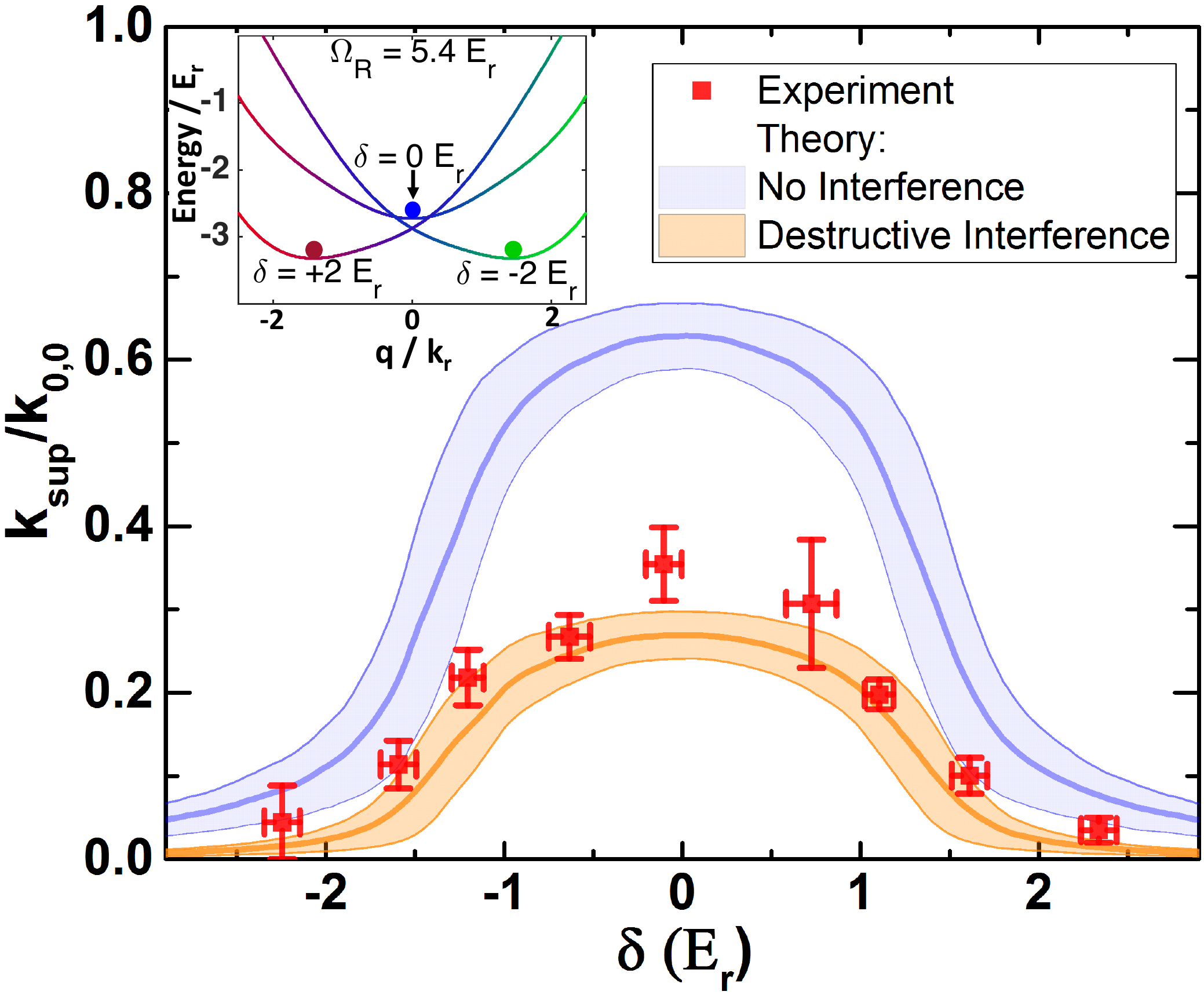}
\caption{Normalized photoassociation rate, $k_{sup}/k_{0,0}$, for BECs at Raman detuning $\delta$ from $-2.5$ to $+2.5\,E_r$ with Raman coupling $\Omega_R=5.4\,\,E_{r}.$ The orange (blue) bands are theoretical predictions with (without) the destructive interference term in Eq. \ref{eq:spinsquare}. The band boundaries reflect one standard deviation of the predicted values given our experimental uncertainties. Inset: dressed band structures for $\Omega_R=5.4\,E_r$ and $\delta = -2,0,$ and $2\, E_r.$ The dots represent BECs prepared at the band minima.}
\label{fig:ksup}
\end{figure}

\red{Describing the entire superposition BEC with one PA rate constant $k_{sup}$ is supported by our observation in Fig. \ref{fig:fraclossfig} that all the bare spin components of the superposition state display identical fractional losses.} In the limit of $\Omega_R\rightarrow0$ with $\delta=0$, $k_\text{sup}/k_\text{0,0}\rightarrow1$ since $C_0\rightarrow1$ and $C_{-1}=C_{+1}\rightarrow0$. However, interestingly, $k_{sup}/k_\text{0,0}\rightarrow0$ for large $\Omega_R$ because $C_0\rightarrow{1}/{\sqrt{2}}$ and $C_{-1}=C_{+1}\rightarrow{1}/{{2}}$ (all the superposition coefficients $C_\text{i}$ can be calculated by diagonalizing Eq. \ref{eq:ham_countprop}). In this case the third term of Eq. \ref{eq:spinsquare} cancels the first two, thus no molecular formation is predicted even with resonant PA light. This happens despite PA being allowed on both channels for associating two $m_f=0$ atoms and associating two $m_f=\pm1$ atoms (see Fig. \ref{fig:Exp_dgm} (a)). This complete destructive interference comes from the opposite CG coefficients ($\pm1/\sqrt{3}$). In Fig. \ref{fig:fraclossfig}, and also later in Fig. \ref{fig:ksup}, the orange (blue) bands are theoretical predictions from Eq. \ref{eq:spinsquare} that show the range of theoretical predictions resulting from our typical experimental uncertainties, including (excluding) the destructive interference effect, the last term of Eq. \ref{eq:spinsquare}. The nearly total suppression of $k_{sup}/k_{0,0}$ at large $\Omega_R$ with $\delta=0\,E_r$ is consistent with the prediction of Eq. \ref{eq:spinsquare} with the destructive interference term. 

We also studied PA on BECs prepared with $\Omega_R = 5.4\,E_r$ and $\delta$ from $-2.5$ to $+2.5\,E_r$. Figure \ref{fig:ksup} shows $k_{sup}/k_{0,0}$ vs $\delta$, measured using square PA pulses with $t_{PA}=5.5$ ms and $I_{PA}=5.7\pm0.2\,\,\text{W}/\text{cm}^2$. The experimental error bars reflect the aggregate uncertainty of $k_{sup}/k_{0,0}$. The inset contains calculated band structures for $\delta=-2,0$ and $+2\,E_r$. \red{Increasing $|\delta|$ beyond $\sim\mp\,2\,E_r$ polarizes the dressed BEC into majority $m_f=\pm\,1$ and collisions between such atoms do not contribute to the $\ket{F=0,m_F=0}$ channel.} This is consistent with our observed $k_{sup}/k_{0,0}\rightarrow0$ with increasing $|\delta|$. This suppression is again consistent with Eq. \ref{eq:spinsquare} because $C_0$ and one of $C_{\pm1}$ vanish, predicting $k_{sup}/k_{0,0}\rightarrow0$.

Atoms in spin-momentum superpositions are novel reactants, and thus PA of such atoms represents a new type of photochemistry. We interpret observing the same fractional loss on all components of a spin-momentum superposition state as an indication that it is the superposition state itself that undergoes PA. Further, the different total fractional loss between BECs of a spin statistical mixture and a spin-momentum superposition state demonstrates a significant modification to the PA process \red{due to the coherent quantum superposition}. Lastly, we interpret the nearly full suppression of $k_{sup}/k_{0,0}$ as resulting from destructive interference between the two out-of-phase pathways ($m_f=0,m_f=0$ and the $m_f=+1,m_f=-1$). Our scattering state simultaneously accesses these two pathways as it couples to the chosen excited molecular state. Thus our observations suggest that scattering states of atoms in quantum superpositions may offer a powerful new approach to coherently control photochemical reactions.


%

\section{Acknowledgements}
We acknowledge helpful conversations with Su-Ju Wang and Abraham Olson. We acknowledge NSF Grant PHY-1708134 and the Purdue University OVPR Research Incentive Grant 206732 for partial support of our research. D.B.B. also acknowledges a Purdue Research Foundation Ph.D. Fellowship. Y.Y. and Q.Z. acknowledge startup funding from Purdue University.
\bibliography{PASOC_Bib}{}
\bibliographystyle{apsrev4-1}

\end{document}